# Building a Virtual HPC Cluster with Auto Scaling by the Docker

Hsi-En Yu and Weicheng Huang
National Center for High-Performance Computing
Hsinchu, Taiwan
yun@narlabs.org.tw, whuang@narlabs.org.tw

*Abstract*—Solving the software dependency issue under the HPC environment has always been a difficult task for both computing system administrators and application scientists. This work would like to tackle the issue by introducing the modern container technology, the Docker, to be specific. By integrating the auto-scaling feature of service discovery with the light-weight virtualization tool, the Docker, the construction of a virtual cluster on top of physical cluster hardware is attempted. Thus, through the isolation of computing environment, a remedy of software dependency of HPC environment is possible.

## I. Introduction

The container technology is a modern virtualization technology at its hype. The Docker is one of the most adopted solutions. On the other hand, the High Performance Computing (HPC) is a mature technology that emphasizes on the parallel performance aiming at number crunching applications. The adoption of modern Docker toolkit onto the HPC environment and applications is one of the hopes for future integration of the resources for computing intensive applications and data intensive applications.

In conventional HPC cluster, we often encounter the conflicts of software versions as well as system characteristics, such as operating system, libraries, kernel, and so on. To resolve such problems, numerous versions of software or libraries are installed and provided on the computing resource. As a consequence, such approach will increase the complexity of the system/middleware of the computing resource and in turn harm the stability of the system. The system administrators have to devote huge efforts to deal with such situations, while users will suffer from the complicated environment for application porting.

The situation will become worse, even become an unsolvable situation, if different versions of OS distribution are encountered. An application that runs well on a distribution of Linux OS, for example the Redhat, is not guaranteed to be executable on another distribution such as the Debian. Usually, the application has to go through a series of modification and recompilation in order to adapt to the new environment imposed by different OS.

To tackle this issue, our work proposed the construction of a virtual HPC cluster via the Docker container. By utilizing the isolation feature of the container technology, a virtual HPC cluster that is independent from other cluster instances can be constructed. The scale of the cluster can also be expanded as needed.

## II. Background

### A. HPC

Nowadays, the HPC clusters are usually running on Unix/Linux-based OS with high-speed interconnect with information exchanged via message passing library such as MPI(Message Passing Interface) and the shared memory directives such as OpenMP for parallel computing jobs. Generally speaking, the bandwidth between compute nodes is the major factor that affects the performance of parallel computing. Usually, the Infiniband or 10GbE-level of network are used as the interconnect. The faster the interconnect is, the better the performance will be.

### B. Docker

The Docker[1] is an open-source project which is an implementation of a lightweight OS level virtualization solution. The core technologies adopted by the Docker are the Linux Container[2] (LXC) and the multi layered file system, the UnionFS[3] (Unification File System). In the LXC era, IT technicians use the cgroup, namespace and chroot to provide the OS level virtualization. However, the architecture is too complicated and difficult to be reproduced or reused.

The Docker is aimed to tackle the issue. Through the encapsulation, IT technicians only need to focus on the management of containers. The Docker engine will take care of the rest. After the completion of an application environment, the system configuration can be packed into the Docker image that can be posted and shared in the Docker Hub[4]. Thus, the efforts can be shared among the community and save lots duplicated development efforts. And in turn, the software build-up and deployment of applications can be speed up. It has become a new trend to deploy software and applications by simply following the standards to encapsulate the software and applications into services and deploy via containers. The Docker engine will handle the interconnect of containers and the services status as well. This kind of system architecture is named as "Microservices" [5].

III. SYSTEM DESIGN

The system architecture of the virtual HPC cluster is shown in Fig. 1. Each container is deployed on separate physical machine. Customized software bridges are built as the containers interconnect. Every container is ready for running the MPI job with the MPI library installed.

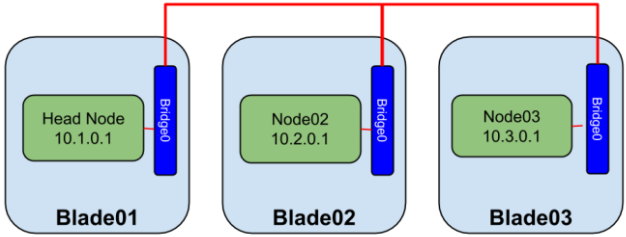

Fig. 1. System Architecture

In order to build the architecture, the following features is needed:

*A. Build the Docker Images for HPC Head/Compute Nodes*

A Dockerfile that includes the requirement of HPC compute nodes, such as the MPI library, OpenSSH server and so on. Once all the components are ready, a Docker images will be built by Docker engine and could be shared to the Docker Hub.

The following figure, in Fig. 2 is an example of Dockerfile for the computing node:

```
[root@blade03 mpi-computenode]# cat Dockerfile
FROM centos:6
MAINTAINER Hsi-En Yu <yun@narlabs.org.tw>

#install software
RUN yum install -y openssh-server openmpi

#install consul-template
ADD consul-template /usr/local/bin/consul-template
ADD consul /usr/local/bin/consul

CMD ["/usr/sbin/sshd", "-D"]

[root@blade03 mpi-computenode]#
```

Fig. 2. The HPC contntainer Dockerfile example

*B. The Communication Between Containers of Cross Nodes*

To provide communication between containers for HPC cluster, the default bridging of containers, the "docker0", is used first. However, the communication range of the default bridge is limited to a single physical machine. In order to provide communication capability across computing nodes for parallel processing, the adjustment of network bridging architecture is needed.

Our work is to build a customized bridge (bridge0) and tell Docker engine to use bridge0 instead, as shown in Fig. 3. The bridge0 binds a physical ethernet interface. This interface is used by the container to connect to the network directly without the need of the NAT translation.

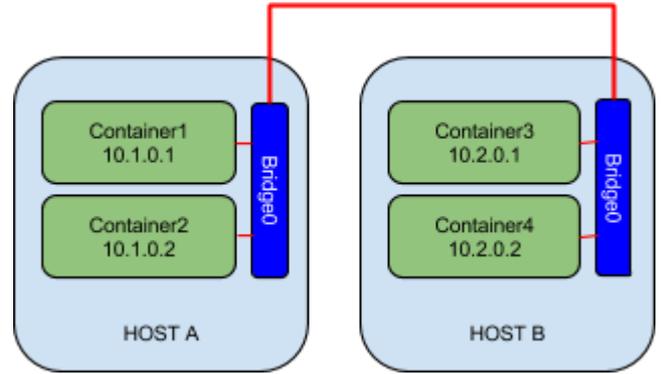

Fig. 3. The communication between containers of cross nodes

*C. Service Discovery and Configuration*

Usually, when a container is initiated, a floating IP is assigned dynamically. To build a cluster, the IP of every node should be acknowledged by all the member nodes in the cluster. Therefore, once a container is initiated, we have to retrieve manually the IP assigned dynamically, thus the cluster hostfile list can be constructed for the use of parallel computing jobs. This will cause problem when initiating tens or even hundreds of containers as cluster member nodes.

The remedy proposed by this work is to use "Consul"[6] to discover the network service. The Consul is a distributed service discovery and configuration service that provides High Availability (HA) mechanism. The Consul agent is built into the container. Thus, all the containers deployed will register to the Consul service automatically. As a consequence, the host file will have the most updated version of IP list assigned to all the containers. All we have to do is to retrieve the computing nodes' list from the headnode and use it as the hostfile list.

IV. IMPLEMETION

In the following session, we will explain how the development work and experiments were carried out. We have adopted three physical servers, namely Blade01, Blade02, and Blade03, to provide infrastructure for the virtual HPC cluster. The hardware specification is shown in the TABLE I.

TABLE I.

| System Model | Dell M620 |
|---|---|
| CPU | Intel(R) Xeon E5-2630 2.30GHz X 2 |
| Memory | 64GB |
| HDD | SAS 146GB 10Krpm |
| Network | 10GbE |

The CentOS 7.1 is used as the OS of all the three physical machines. The CentOS 7 is a Linux distribution with native support to the Docker. The rest of major software versions are shown in the TABLE II.

TABLE II.

| Physical Machine OS | CentOS 7.1.1503 x64 |
|---|---|
| Docker Engine | Docker version 1.5.0-dev build fc0329b/1.5.0[a] |
| Consul | Consul v0.5.2 |
| Container OS | CentOS 6.7 |
| MPI Library | OpenMPI (the latest version in CentOS 6.7) |

[a.] Redhat official package

As shown in the system architecture, the Fig. 4, a container "Head" is activated on the physical node Blade01. The "Head" serves as the head node of the virtual cluster. In addition, 2 containers, the node02 and node03, are activated on the Blade02 and Blade03.

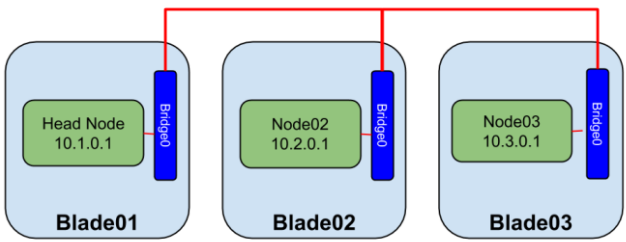

Fig. 4. System Architecture of virtual HPC cluster

In order to provide Consul service, a distributed Consul service is setup outside of the system and the Consul agent is installed on the HPC Docker Images. Once the "Node02" and "Node03" are deployed, the entries will be log on to the Consul service declaring HPC service for the two containers.

Meanwhile, the head node will retrieve the dynamical IP list from the Consul server through the Consul-template. Consul-template[7] is a software project for generic template rendering and notifications with Consul service. The retrieved IP list will be used to construct the hostfile list, which will be used to provide service to the MPI parallel jobs, as shown in Fig. 5. Users do not have to worry about the hostfile at all. If more computing power is needed, all we need to do is to power up more physical machines and deploy new HPC containers on those machines. The new HPC containers will register themselves to the Consul automatically and become part of the computing cluster.

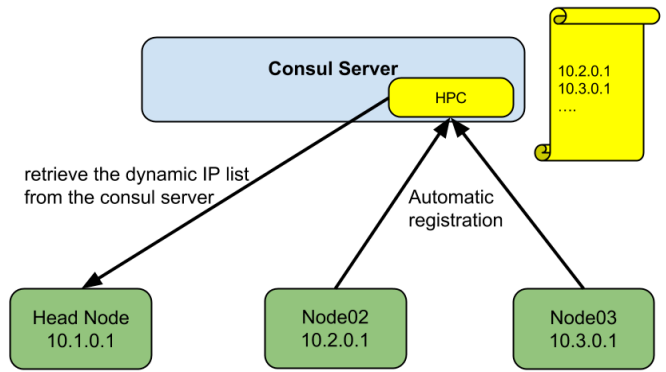

Fig. 5. The service discovey scheme of HPC by Consul

The present implementation is shown in following screenshots. The Fig. 6 shows all the three servers with Docker containers running as cluster nodes. The auto registration of containers to the Consul are displayed in Fig. 7, followed by the Fig. 8 in which the execution of a MPI job is shown.

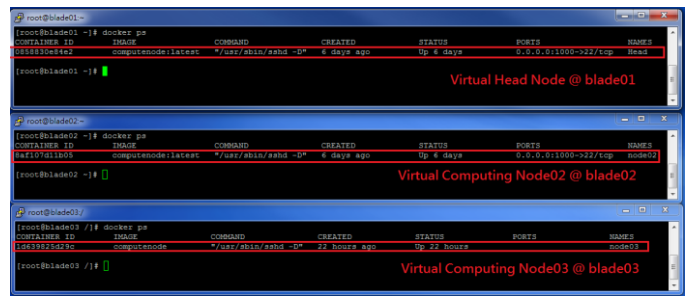

Fig. 6. Three containers are running on separate physical machine.

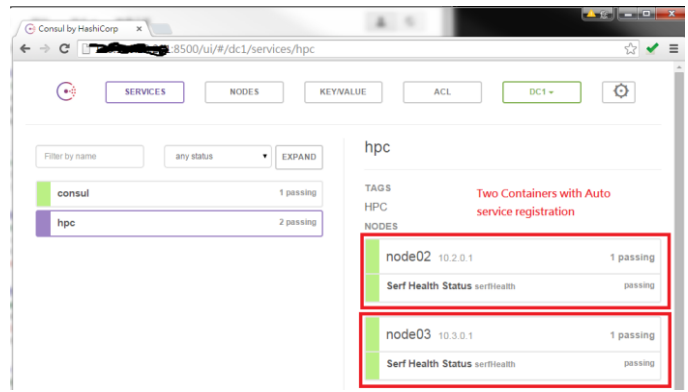

Fig. 7. The containers register themselves to the Consul service

Fig. 8. Execution of a 16-domain MPI job on the virtual HPC cluster with 2 containers.

## V. CONCLUSION

Through the adoption of the container technology, a prototype of virtual HPC cluster with auto scaling is demonstrated. Our work shows the feasibility of virtual HPC cluster. It is also possible to use this prototype to build a customized virtual HPC cluster above the HPC infrastructure for specific applications. However, performance is always an important issue for the parallel computing. It is our intention to further investigate the performance of this prototype, including the influence of the interconnect between HPC containers. In addition, to facilitate the usage of the approach, a more powerful container management tool, such as Docker Swarm or Kubernetes, as well as a user friendly GUI are needed in order to move toward the goal of HPC-as-a-Service (HaaS).